\documentclass[aps,showpacs,twocolumn,superscriptaddress]{revtex4}

\usepackage{graphicx}
\usepackage{dcolumn}

\begin{document} 

\title{Comment on ``Lattice determination of $\Sigma$-$\Lambda$ mixing"} 

\author{A.~Gal}\affiliation{Racah Institute of Physics, 
The Hebrew University, Jerusalem 91904, Israel} 

\begin{abstract} 

A recent lattice QCD (LQCD) calculation of $\Sigma$-$\Lambda$ mixing by the 
QCDSF-UKQCD Collaboration [Phys. Rev. D 91, 074512 (2015)] finds a mixing 
angle about half of that found from the Dalitz-von Hippel (DvH) flavor SU(3) 
mass formula which relates the $\Sigma$-$\Lambda$ mixing matrix element to 
known octet baryon mass differences and which has been used widely to evaluate 
charge symmetry breaking effects in $\Lambda$ hypernuclei. We show that the 
LQCD-calculated $\Sigma$-$\Lambda$ mixing matrix element and octet baryon 
masses satisfy the DvH mass formula, concluding thereby that a good LQCD 
evaluation of $\Sigma$-$\Lambda$ mixing requires an equally good reproduction 
of octet baryon mass differences which is yet to be demonstrated. 

\end{abstract} 

\pacs{12.38.Gc, 11.30.Hv, 12.39.Jh, 21.80.+a} 

\maketitle 


Back in 1964, Dalitz and von Hippel (DvH) observed that the lowest-lying 
neutral hyperons $\Lambda^0$ and $\Sigma^0$ have mixed isospin composition 
in terms of the SU(3) octet pure isospin $\Lambda$ ($I=0$) and $\Sigma$ 
($I=1$) baryons, 
\begin{equation} 
\Lambda^0 = \Lambda\;\cos\alpha - \Sigma\;\sin\alpha,\;\;
\Sigma^0 = \Lambda\;\sin\alpha + \Sigma\;\cos\alpha, 
\label{eq:mix1} 
\end{equation} 
where the $\Sigma$-$\Lambda$ mixing angle $\alpha$ is given by
\begin{equation} 
\tan\alpha= \langle\Sigma|\delta M|\Lambda\rangle/(M_{\Sigma}-M_{\Lambda}), 
\label{eq:mix2} 
\end{equation} 
with mixing mass $\delta M$ matrix element related 
to isospin-breaking mass differences within the baryon octet \cite{DvH64}: 
\begin{equation} 
\langle\Sigma|\delta M|\Lambda\rangle=\frac{1}{\sqrt 3}
[(M_{\Sigma^0}-M_{\Sigma^+})-(M_n-M_p)].   
\label{eq:deltaM1} 
\end{equation}
Another expression was derived in the quark model in terms of isospin-breaking 
mass differences within the baryon {\bf 56} SU(6) multiplet \cite{GS67}: 
\begin{equation} 
\langle\Sigma|\delta M|\Lambda\rangle=\frac{1}{2\sqrt 3}
[(M_{\Xi^-}-M_{\Xi^0})-(M_{\Xi^{\ast -}}-M_{\Xi^{\ast 0}})]. 
\label{eq:deltaM2} 
\end{equation}
The right-hand sides of these two mass formulae agree with each other within 
the PDG quoted errors \cite{PDG14}. Here we use the DvH SU(3) mass formula 
(\ref{eq:deltaM1}) owing to its smaller errors. Its left-hand side 
$\Sigma$-$\Lambda$ mixing output has been widely used in $\Lambda$ 
hypernuclear evaluations of charge symmetry breaking (CSB), see 
Refs.~\cite{MAMI15,Gal15} and references therein for the latest 
experimental and theoretical CSB state of the art in hypernuclei. 

\begin{table}[ht] 
\caption{$\langle\Sigma|\delta M|\Lambda\rangle$, Eq.~(\ref{eq:deltaM1}) 
in MeV, and the resulting $\Sigma$-$\Lambda$ mixing angle $\alpha$, 
Eq.~(\ref{eq:mix2}), for PDG and LQCD input masses. } 
\label{tab:mix} 
\begin{tabular}{lcc} 
\hline \hline 
Input masses & $\langle\Sigma|\delta M|\Lambda\rangle$ & 
$\alpha$ \\ 
\hline 
PDG \cite{PDG14} & 1.14(05) & 0.015(1) \\ 
LQCD \cite{QCD15}~~~~~& 0.52(23) & 0.007(3) \\ 
\hline\hline 
\end{tabular} 
\end{table} 

A recent LQCD calculation \cite{QCD15} reporting on a pure QCD calculation, 
short of QED effects, finds a $\Sigma$-$\Lambda$ mixing angle $\alpha_{LQCD}=
0.006(3)$. This is less than half the mixing angle $\alpha_{\rm DvH}=0.015(1)$ 
obtained by using PDG mass values in the DvH mass formula (\ref{eq:deltaM1}), 
as listed in Table~\ref{tab:mix}. We note that using LQCD-calculated input 
masses in the DvH mass formula leads to the value $\alpha = 0.007(3)$, also 
listed in the table, agreeing nicely with the calculated value $\alpha_{LQCD}=
0.006(3)$. This means that the DvH SU(3) mass formula (\ref{eq:deltaM1}) is 
satisfied by the LQCD calculation \cite{QCD15}, although at least one of its 
mass-difference terms misses appreciably the corresponding PDG value, or the 
value naively expected upon disregarding electomagnetic (em) effects. 
We note that the dominant quark-quark Coulomb interactions cancel out in 
the mass differences appearing in the DvH mass formula (\ref{eq:deltaM1}), 
and the magnetic interactions are an order of magnitude weaker, resulting 
in a tiny em contribution of order 0.1~MeV \cite{Isgur80} to its left-hand 
side. It is, therefore, unclear whether or not the discrepancy pointed out 
in the present Comment will get resolved upon including em effects in this 
particular LQCD calculation. 

We note, in this connection, that the baryon octet masses calculated in the 
LQCD work \cite{QCD15} satisfy perfectly also the Coleman-Glashow (CG) mass 
formula \cite{CG61}, 
\begin{equation} 
(M_n-M_p)-(M_{\Sigma^-}-M_{\Sigma^+})+(M_{\Xi^-}-M_{\Xi^0})=0, 
\label{eq:CG} 
\end{equation} 
although none of its three CSB-like mass differences comes satisfactorily 
close to the corresponding PDG value and this discrepancy persists for some 
of these mass differences upon disregarding em effects. These mass differences 
have been calculated recently by the BMW Collaboration \cite{BMW15}, 
demonstrating good agreement with the PDG values by including em effects 
upon using a given convention to define the distinction between QCD and QED 
contributions. Missing unfortunately in such calculations is the ($\Sigma^0-
\Sigma^+$) mass difference that enters the DvH formula (\ref{eq:deltaM1}).

Given the correlation demonstrated above between the $\Sigma$-$\Lambda$ 
mixing matrix element and the LQCD-calculated octet baryon mass differences 
appearing in the DvH mass formula, as well as the correlation between 
those octet baryon mass differences of which the CG mass formula consists, 
we question the reliability of the $\Sigma$-$\Lambda$ mixing matrix element 
and the associated $\Sigma$-$\Lambda$ mixing angle derived in the recent LQCD 
work \cite{QCD15}, so long as the related PDG baryon mass differences are 
still far from being reproduced to a good accuracy. A good LQCD evaluation 
of $\Sigma$-$\Lambda$ mixing requires an equally good reproduction of octet 
baryon mass differences which is yet to be demonstrated. 
Finally, we note that accepting the relatively small LQCD $\Sigma$-$\Lambda$ 
mixing angle would be difficult to reconcile with the large CSB splitting 
observed and discussed in the $A=4$ charge symmetric hypernuclei 
$_{\Lambda}^4$H--$_{\Lambda}^4$He \cite{MAMI15,Gal15}.

\end{document}